\documentclass[11pt]{article}
\usepackage[a4paper, left=2.7cm, right=2.7cm, top=2.8cm, bottom=2.8cm]{geometry}
\usepackage{amsmath,amssymb,amsthm}
\usepackage{ulem}
\usepackage{url}
\usepackage{booktabs}
\usepackage{xcolor}
\usepackage{natbib}
\usepackage{setspace}
\usepackage{microtype}
\usepackage{authblk}
\usepackage{hyperref}

\newcommand{\tr}{{\prime}}

\newcommand{\calM}{\mathcal{M}}
\newcommand{\Yobs}{Y^{\mathrm{obs}}}
\newcommand{\yobs}{y^{\mathrm{obs}}}

\newcommand{\MM}{\underline{M}}
\newcommand{\mm}{\underline{m}}

\newcommand{\XX}{\underline{X}}
\newcommand{\xx}{\underline{x}}
 
\newcommand{\Mobs}{M^{\mathrm{obs}}}

\newcommand{\Rreg}{RR^{\mathrm{reg}}}
\newcommand{\MMobs}{\underline{M}^{\mathrm{obs}}}

\newcommand{\ci}{\!\perp \! \! \! \perp\!}

\newenvironment{keywords}
{\par\noindent\textbf{Keywords: }\ignorespaces}
{\par}

\newtheorem{assumption}{Assumption}

\title{Regression-based approach for natural direct and indirect relative risk in case of multiple mediators}

\author{
Monia Lupparelli, Arianna Nuti, Giovanni Maria Marchetti, Alessandra Mattei
\thanks{E-mail addresses:
\href{mailto:monia.lupparelli@unifi.it}{monia.lupparelli@unifi.it},
\href{mailto:arianna.nuti@unifi.it}{arianna.nuti@unifi.it},
\href{mailto:giovanni.marchetti@unifi.it}{giovanni.marchetti@unifi.it}, 
\href{mailto:alessandra.mattei@unifi.it}{alessandra.mattei@unifi.it}.}
\\
\small Department of Statistics, Computer Science, Applications, University of Florence, Florence, Italy\\
}

\date{ }

\begin{document}

\maketitle

\begin{abstract}
Mediation analysis investigates whether part of the treatment effect is channelled through one or more mediators along the causal pathway between the treatment and the primary outcome.  However, the presence of multiple, potentially dependent mediators raises substantial challenges, particularly when the outcome is binary, the mediators are measured on different scales, and interactions are present. 
We consider on causal mediation analysis with a binary treatment and a binary outcome, defining natural direct, indirect, and total effects on the relative-risk scale, thereby avoiding the interpretational difficulties associated with non-collapsible effect measures. 
Under a sequential ignorability assumption, we develop a unified regression-based framework that accommodates multiple continuous, binary, or mixed mediators. 
The proposed framework accounts for dependence among mediators and allows for both exposure–mediator and mediator–mediator interactions. We derive closed-form expressions for the causal effects across the different mediator settings and develop a likelihood-based inference procedure for estimating the causal effects and quantifying their uncertainty.
The methodology is illustrated through two empirical applications.
\end{abstract}


\section{Introduction}
\label{s:intro}

Casual mediation analysis focuses on understanding  the mechanisms through which a treatment affects an outcome by investigating the role of intermediate variables, known as mediators, in the treatment--outcome relationship.
When part of the treatment effect is channelled through mediators, the primary objective is to disentangle indirect effects, operating through the mediators of interest, from direct effects, operating through alternative pathways \citep{RobinsGreenland1992, Pearl2001, Imaietal:2010, VanderWeele2015}.
 
When the outcome is binary, causal effects may be defined on the risk difference, odds ratio, or risk ratio scale, which  can substantially differ for non-rare outcomes \citep{hernan2025causal}. We focus on natural direct and indirect effects defined on the relative risk scale in the presence of multiple mediators. In particular, we consider a joint natural indirect relative risk capturing the effect channelled through the mediators as a whole, and a natural direct relative risk representing the effect operating outside all mediating pathways.
The relative risk scale is widely used in epidemiologic and biomedical studies because it has a straightforward probabilistic interpretation as the ratio of outcome risks under different exposure levels. 
Furthermore, unlike the odds ratio, the relative risk is a collapsible measure \citep{green-et-al:1999, forastiere2025selecting}.
Despite these advantages, regression-based inference for natural direct and indirect relative risks remains challenging, mainly in the presence of multiple mediators. The need to integrate over the mediator distribution, together with the complexity of multiple mediators, further complicates estimation and inference. This task depends critically on their measurement scales, their dependence structure, and the interactions included in the outcome model. 

Under a sequential ignorability assumption, we propose a regression-based framework for inference on natural direct and indirect relative risks with a binary treatment, a binary outcome, and multiple contemporaneous mediators.
The main contribution is to provide a unified formulation accommodating continuous, binary, and mixed mediators, while allowing both exposure--mediator and mediator--mediator interactions.
The framework combines a log-link regression model for the binary outcome with a joint model for the mediators: a multivariate log-mean regression model \citep{lup-rov:2017} for binary mediators, a multivariate linear regression model for continuous mediators, and a conditional Gaussian specification for mixed mediators. Under correct model specification, the natural direct and indirect relative risks admit closed-form analytical expressions as functions of the model parameters on the log-relative risk scale. The resulting natural direct and indirect relative risks provide a multiplicative decomposition of the total effect.

Beyond this unified formulation, the paper makes two specific methodological contributions. For continuous mediators, we derive closed-form expressions for natural effects when mediator--mediator interactions are included in the outcome model. Although such interactions have been considered in regression-based mediation analysis \citep{VanderWeeleVansteelandt2014}, their inclusion requires integrating nonlinear functions of the joint mediator distribution, and explicit analytical expressions for the resulting effects have not been provided yet. For binary mediators, we derive natural direct and indirect relative risks under a joint model that accommodates both dependence and mediator--mediator interactions. 
In this setting, the marginal and joint mediator models cannot be specified independently, leading to a parameter space shaped by complex non-linear constraints that substantially complicate inference.
We explicitly address these constraints within the proposed regression framework by developing likelihood-based inference for the multivariate log-mean regression model. Building on the maximization procedure proposed by \citet{eva-for:2013} for a broad class of generalized linear regression models, we implement an algorithm adapted to the log-mean link that preserves the parameter-space constraints and yields estimates corresponding to a coherent joint mediator distribution.

The  methodology is applied to two empirical applications using data  from the National Health and Nutrition Examination Survey and the Interdisciplinary Project for the Optimization of Separation trajectories study, respectively.

\section{Existing Approaches for Multiple Mediators}
\label{s:review}
 
Various methodological approaches extending mediation analysis to the multiple-mediator setting have been proposed in the literature, each relying on different assumptions regarding the dependence structure of mediators and tailored to distinct causal questions.

In some studies, a particular mediator is of primary interest, whereas other post-treatment variables are treated as exposure-induced mediator--outcome confounders. The objective is then a  decomposition of the total effect into an indirect effect  through the mediator of interest and a direct effect encompassing all pathways that do not involve that mediator
\cite[e.g.][]{vanderweele2014effect, vanderweele2014sensitivity, tchetgen2012semiparametric}.  

When all mediators are of substantive interest, different causal estimands may be considered depending on the scientific question and on the assumptions made about the causal relationships among the mediators.
Under the assumption that mediators do not causally influence one another, mediator-specific indirect effects can be defined \citep{mackinnon2000contrasts, preacher2008asymptotic}. 
When mediators may causally affect one another, path-specific effects can be used to quantify effects transmitted along particular causal pathways. These effects are defined relative to a directed acyclic graph (DAG) that specifies the causal ordering among mediators and may be used either to isolate effects operating through selected pathways \citep{AvinShpitserPearl:2005, albert2011generalized} or to decompose the total effect into pathway-specific components \citep{Daniel_et_al:2015}. Because their definition depends on the underlying causal graph, assumptions about mediator ordering are crucial. \citet{ImaiYamamoto2013} examine the sensitivity of mediation analyses to alternative assumptions about causal relationships among mediators.

When the causal ordering among the mediators is unknown or is not  of scientific interest, interventional direct and indirect effects provide an alternative to path-specific natural effects, as they do not require specification of the causal ordering among mediators\citep{VansteelandtDaniel2017, loh2022nonlinear}.
Alternatively, one may focus on the effects mediated jointly through the full set of mediators \citep{WangEtAl:2013, VanderWeeleVansteelandt2014, nguyen2016causal}.
We contribute to  this line of research.  
Unlike mediator-specific, path-specific, and interventional decompositions, this approach treats the mediators as a joint mediating mechanism and does not attempt to attribute the indirect effect to individual mediators or causal pathways.

Related works include \citet{WangEtAl:2013}, who use the mediation formula to estimate the joint natural effect of multiple mediators of different types on a binary outcome, and \citet{nguyen2016causal}, who consider multiple continuous or ordinal mediators within a structural equation modelling framework. \citet{VanderWeeleVansteelandt2014} propose regression- and weighting-based approaches for assessing the effect mediated jointly through multiple binary or continuous mediators, allowing exposure--mediator interactions and certain forms of mediator--mediator interaction. More recently, \citet{raggi2023path} develop a regression framework for a binary variable setting where direct and indirect effects are obtained analytically from the model parameters. These effects are defined on the log-odds scale, and the decomposition of the total effect generally requires an additional residual component.

Our method contributes to this literature by developing a unified generalized multivariate  regression framework for natural direct and joint indirect effects on the relative-risk scale. The proposed setting yields closed-form expressions applicable to binary, continuous, and mixed mediators, encompassing both exposure–mediator and mediator–mediator interactions. 

\section{Causal Estimands}
\label{s:assumptions}
Let $Y$ be a binary outcome, $W$ a binary treatment and let $\MM=\left(M_1, \ldots, M_k, \ldots, M_K\right)^\tr$ denote a vector of $K$ contemporaneous mediators, post-treatment intermediate variables lining in the causal pathway between the treatment and the outcome.
The outcome and the treatment take value $y,w \in \{0,1\}$, respectively, and the vector of the mediators, $\MM$,  takes on value $\mm \in {\cal{M}} \subseteq R^K$, e.g., ${\cal{M}}=\{0,1\}^K$ whether it is a vector of binary variables and ${\cal{M}}$ is an continuous subset in  $R^K$, e.g., $[m_l, m_u]^K$ with $m_l <m_u \in R$,  whether it consists of continuous variables.  $\XX$ is a $p-$dimensional vector of  covariates taking value $\xx \in R^p$. 

Under the Stable Unit Treatment Value Assumption \cite[SUTVA,][]{Rubin:1990}, let  $\MM(w)=\left(M_1(w), \ldots, M_k(w), \ldots, M_K(w)\right)^\tr$ and $Y(w)$ denote the potential outcomes for the $K$ mediators and the primary endpoint, respectively: they are the values of $\MM$ and $Y$ under  treatment $w$, $w=0,1$. Let $W$, $\MMobs=\left(\Mobs_1, \ldots, \Mobs_k, \ldots,\Mobs_K\right)^\tr$ and $\Yobs$ denote the observed values of the treatment, the mediators and the outcome: $\MMobs=W \,\MM(1) + (1-W)\,\MM(0)$ with
$\Mobs_{k}=M_{k}(W)=W \, M_{k}(1) + (1-W)\, M_{k}(0)$ and $\Yobs=Y(W)=W \, Y(1) + (1-W)\, Y(0)$.

We are interested in decomposing the total treatment effect  of  $W$ on the outcome $Y$,  into a joint indirect relative risk, representing the effect mediated collectively by $\MM$, and a direct relative risk, representing the effect transmitted via pathways  bypassing all mediators in $\MM$.

In order to formally define the casual effects of interest we introduce  potential outcomes of the form  $Y(w,\mm)$, $\mm \in \calM$ and $Y(w,\MM(w^\ast))$, $w,w^\ast \in \{0,1\}$: $Y(w,\mm)$ would be the value of the outcome $Y$ if the treatment were set to the level $w$ and the multivariate mediator $\MM$ were set to the value $\mm$; and $Y(w,\MM(w^\ast))$ would be the value of the outcome $Y$ if the treatment were set to the level $w$ and all the mediators $\MM$ were set to the values they would have taken if the treatment had been set to an alternative level, $w^\ast$. These types of potential outcomes are well defined under an augmented SUTVA, requiring that outcome values for a unit do depend on neither treatments nor values for the mediators of other units, and that there is ``no multiple versions'' of treatment and  intermediates \citep{Kim_et_al:2019}.

We define causal effects conditional on the covariates. The corresponding marginal causal effects can be derived by averaging over the empirical distribution of the covariates. The causal relative risk conditional on covariates level $\XX=\xx$ is
\begin{equation}
RR_{Y|\xx}= \frac{P(Y(1)=1\mid \XX=\xx)}{P(Y(0)=1 \mid \XX=\xx)}=\frac{P(Y(1, \MM(1))=1 \mid \XX=\xx)}{P(Y(0, \MM(0))=1 \mid \XX=\xx)}, \notag
\end{equation}
which can be decomposed into the product of a `natural' direct relative risk and  a joint `natural' indirect relative risk: $RR_{Y|\xx}=RR^{\;dir}_{Y|\xx} \times RR^{\;ind}_{Y|\xx}\;$: 
\begin{equation}
RR^{\;dir}_{Y|\xx}=\frac{P(Y(1,\MM(0))=1|\XX=\xx)}{P(Y(0,\MM(0))=1|\XX=\xx)}, \notag \quad \quad RR^{\;ind}_{Y|\xx}=\frac{P(Y(1,\MM(1))=1|\XX=\xx)}{P(Y(1,\MM(0))=1|\XX=\xx)} \notag.
\end{equation}

For the identification and estimation of direct and indirect relative risks we invoke two assumptions on the treatment assignment mechanism and the   mediators-outcome relationship. 
\begin{assumption} (Ignorability of the treatment). \label{ass:1}
	$$\{Y(w, \mm), \MM(w^\ast)\} \ci W  \mid \XX \qquad  \hbox{for all } w,w^\ast \in \{0,1\}, \mm \in {\cal{M}}
	$$
\end{assumption}
\begin{assumption} (Ignorability of the mediators). \label{ass:2}
	$$  Y(w, \mm) \ci \MM(w^\ast) \mid W=w^\ast, \XX \qquad  \hbox{for all } w,w^\ast \in \{0,1\}, \mm \in {\cal{M}} 
	$$
\end{assumption}  
Assumptions \ref{ass:1} and \ref{ass:2} jointly generalize  the   \textit{sequential ignorability} assumption usually invoked in mediation analysis with a single mediator \cite[e.g.,][]{FMD:2018,Imaietal:2010}. \cite{VanderWeeleVansteelandt2014} provide an alternative but similar set of identifying assumptions for natural direct and indirect effects with multiple mediators. 
Ignorability of the treatment states that the  treatment assignment mechnism is ignorable conditional on the observed covariates and holds by design in randomized experiments.  Ignorability of the multiple mediator  states that the mediators are ignorable given the observed treatment and covariates.
Under Assumptions~\ref{ass:1} and \ref{ass:2}, the mediation formula holds \citep{Pearl2001}:
\begin{eqnarray}
	\lefteqn{P\left(Y(w, \MM(w^\ast))=1 \mid \XX=\xx\right)=} \nonumber\\&& 
	\int_{\mm \in {\cal{M}}}P(\Yobs=1\mid \MMobs=\mm,W=w,\XX=\xx)f_{\MMobs\mid w^\ast,\xx}(\mm\mid w^\ast,\xx) \nu(d\mm) \label{eq:medformula}
\end{eqnarray}
where
$f_{\MMobs\mid w^\ast,\xx}(\mm \mid w^\ast,\xx)$ is the  joint probability density/mass function for the random variable $\MMobs\mid \{W=w^\ast,\XX=\xx\}$, with $w^\ast \in \{0,1\}$ and $\xx \in R^p$, and  $\nu$ is the Lebesgue measure if 
$\MM$ is  continuous   and the counting measure if $\MM$ is discrete (see the Supplementary Material). 

We conduct mediation analysis using the mediation formula in Equation~\eqref{eq:medformula} together with a recursive regression-based approach. 
We assume that the observed data arise from regression models without exposure--mediator interactions; the extension to models including multiplicative exposure--mediator interactions is straightforward and is described in Sections A and B of the Supplementary Material. Under Assumptions \ref{ass:1} and \ref{ass:2}, the mediation formula identifies the natural direct and indirect relative risks as functions of the associational parameters governing the distributions of the observed mediators and outcome. Thus, these causal effect measures can be estimated by fitting standard (multivariate) regression models to the observed data and evaluating the corresponding functions of the estimated regression parameters.

\section{A Recursive Regression Framework with a Single Mediator}
\label{s:single_med}

We first develop the proposed recursive regression framework for a single mediator, $\MM= M_1 \equiv M$ taking on values $m \in {\cal{M}} \subseteq R$.

\subsection{Single Binary Mediator}
\label{s:single_binmed}
We consider a recursive regression framework based on the log-link function for the binary random vector $(\Mobs,\Yobs \mid W, \XX)$ :
\begin{eqnarray}\label{eq:1}
\log P(\Mobs=1 \mid W,\XX)&=&  \beta_\emptyset + \beta_W W + \beta_X^\tr \XX\\
\log P(\Yobs=1\mid \Mobs,W,\XX)& =&  \alpha_\emptyset  + \alpha_W W + \alpha_M \Mobs + \alpha_X^\tr \XX. \label{eq:2}
\end{eqnarray}
Regression coefficients in Equations \eqref{eq:1} and \eqref{eq:2} correspond  to the logarithm of   conditional relative risks: For any $\xx \in R^p$, $m \in {\cal{M}}=\{0,1\}$ and $w \in  \{0,1\}$,
$$
\begin{array}{ccccl}
\beta_W &\!\!\!=\!\!\!& \log \Rreg_{M \mid W.\xx}&\!\!\! =\!\!\!& \log \dfrac{P(\Mobs=1\mid W=1,\XX=\xx)}{P(\Mobs=1\mid W=0,\XX=\xx)},\\
\\
\alpha_W &\!\!\!=\!\!\!&\log  \Rreg_{Y \mid W.m,\xx}& \!\!\!=\!\!\!& \log \dfrac{P(\Yobs=1\mid W=1,\Mobs=m,\XX=\xx)}{P(\Yobs=1\mid W=0,\Mobs=m,\XX=\xx)};\\
\vspace{-0.2cm}\\
\alpha_M &\!\!\!=\!\!\!&\log  \Rreg_{Y \mid M.w,\xx}&\!\!\! =\!\!\!& \log \dfrac{P(\Yobs=1\mid W=w,\Mobs=1,\XX=\xx)}{P(\Yobs=1\mid W=w,\Mobs=0,\XX=\xx)}. 
\end{array}
$$
The $p-$dimensional parameter vectors $\alpha_X$ and $\beta_X$ represent the effect of the individual covariate $\XX$ on the distribution of $Y$ and of $M$, respectively.

Under Assumptions~\ref{ass:1} and \ref{ass:2}, and under correct specification of
the regression models in   \eqref{eq:1}-\eqref{eq:2}, the mediation formula implies that the  natural direct and indirect relative risks are given by 
\begin{eqnarray*}
RR^{\;dir}_{Y\mid \xx}&=&\frac{[1-e^{(\beta_{\emptyset}+\beta_X \xx)}]e^{\alpha_W}+e^{(\beta_{\emptyset}+\beta_X \xx)}e^{(\alpha_W + \alpha_M)}}{1-e^{(\beta_{\emptyset}+\beta_X \xx)}+e^{(\beta_{\emptyset}+\beta_X \xx)}e^{\alpha_M}}= e^{\alpha_W},\\
RR^{\;ind}_{Y|\xx} &=&\frac{[1-e^{(\beta_{\emptyset}+\beta_W+\beta_X^\tr \xx)}]e^{(\alpha_\emptyset + \alpha_W+\alpha_X^\tr \xx)}+e^{(\beta_{\emptyset}+\beta_W+\beta_X^\tr \xx)}e^{(\alpha_\emptyset +\alpha_W + \alpha_M+\alpha_X^\tr \xx)}}{[1-e^{(\beta_{\emptyset}+\beta_X^\tr \xx)}]e^{(\alpha_\emptyset + \alpha_W+\alpha_X^\tr \xx)}+e^{(\beta_{\emptyset}+\beta_X^\tr \xx)}e^{(\alpha_\emptyset + \alpha_W+\alpha_M+\alpha_X^\tr \xx)}}\\
&=&\frac{[1-e^{(\beta_{\emptyset}+\beta_W+\beta_X^\tr \xx)}]+e^{(\beta_{\emptyset}+\beta_W+\beta_X^\tr \xx)}e^{\alpha_M}}{[1-e^{(\beta_{\emptyset}+\beta_X^\tr  \xx)}]+e^{(\beta_{\emptyset}+\beta_X^\tr \xx)}e^{\alpha_M}}.
\end{eqnarray*}
It is worth noticing that, under the conventional assumption that $\Rreg_{Y\mid M.w,\xx}=1$ for $\Mobs=0$, the natural indirect relative risk can be written as ratio of weighted averages of conditional relative risks of the mediator $\Mobs$  on the outcome, $\Yobs$,   with weights given by 
  the conditional probabilities $P(\Mobs=m\mid W=w, \XX)$, $w=0,1$:
$$
RR^{\;ind}_{Y\mid \xx}=\dfrac{\sum_{m =0}^1 \Rreg_{Y\mid M.w,\xx}P(\Mobs=m\mid W=1, \XX=\xx)}{\sum_{m =0}^1 \Rreg_{Y\mid M.w,\xx}P(\Mobs=m\mid W=0, \XX=\xx)}.
$$
See Section A of the Supplementary Material for details. The exposure-mediator interaction  can be also considered by including the interaction term $W \times \Mobs$ in Equation \eqref{eq:2}. The resulting formula for natural direct and indirect relative risks are provided in Section A of the Supplementary Material.

\subsection{Single Continuous Mediator}
\label{s:single_conmed}
In studies where the mediator is continuous, we assume that   $M(w)\mid \XX \sim N\left(E[M(w)\mid \XX], \sigma^2\right)$ and  we use the following   recursive regression framework:
\begin{eqnarray}\label{eq:3}
	E[\Mobs\mid W,\XX]&=&    \gamma_\emptyset + \gamma_W W + \gamma_X^\tr \XX\\
	\log P(\Yobs=1 \mid W, \Mobs,\XX) &=&  \alpha_\emptyset + \alpha_W W + \alpha_M \Mobs +  \alpha_X^\tr \XX. \label{eq:4}
\end{eqnarray}
Under Assumptions~\ref{ass:1} and \ref{ass:2}, assuming correct specification of the regression models  in \eqref{eq:3}-\eqref{eq:4},  using the mediation formula  the  natural direct and indirect relative risks are   given by 
\begin{eqnarray*}
RR^{\; dir}_{Y \mid \xx}&=&\dfrac{e
^{(\alpha_\emptyset + \alpha_W  + \alpha_X^\tr \xx)}e^{\big[(\gamma_\emptyset + \gamma_X^\tr \xx)\alpha_M + \sigma^2 \frac{\alpha^2_M}{2}\big]}}{e^{(\alpha_\emptyset + \alpha_X^\tr \xx)}e^{\big[(\gamma_\emptyset  + \gamma_X^\tr \xx)\alpha_M + \sigma^2\frac{\alpha^2_M}{2}\big]}}= e^{\alpha_W},\\
RR^{\; ind}_{Y\mid \xx}&=&\dfrac{e
^{(\alpha_\emptyset + \alpha_W  + \alpha_X^\tr \xx)}e^{\big[(\gamma_\emptyset +\gamma_W  + \gamma_X^\tr \xx)\alpha_M + \sigma^2\frac{\alpha^2_M}{2}\big]}}{e
^{(\alpha_\emptyset + \alpha_W  + \alpha_X^\tr \xx)}e^{\big[(\gamma_\emptyset  + \gamma_X^\tr \xx)\alpha_M + \sigma^2 \frac{\alpha^2_M}{2}\big]}}=e^{ \gamma_W \alpha_M}.
\end{eqnarray*}
See Section A of the Supplementary Material for details.
In case of a continuous mediator, both the natural direct and indirect effect are independent of the covariate set under the  regression models in \eqref{eq:3} and \eqref{eq:4}. We  remark that the indirect effect specified on the odds ratio scale through logistic regression models, is not invariant with respect to the covariates, even for the case of continuous mediators; see \citet{VanderWeeleVansteelandt2010}.
Formulas for the natural direct and indirect effects which also account for the exposure-mediator interaction  $W \times \Mobs$ in Equation \eqref{eq:4} are provided in Section A of the Supplementary Material.

\section{A Recursive Regression Framework 
with Multiple Mediators}
\label{s:multiple_med}

We  generalize the proposed recursive regression framework to the case of multiple contemporaneous mediators. Let $\MM =(M_k)_{k \in K}$  denote a vector of potentially dependent mediators, where $K$ is a finite index set with $|K|>1$. We derive expressions for the natural direct and indirect relative risks associated with the joint mediating mechanism represented by $\MM$.

\subsection{Multiple Binary Mediators}
\label{s:multiple_binmed}

For any nonempty subset $D \subseteq \{1, \ldots, K\}$, let $M_D=\prod_{k \in D} M_k$ be the product-mediator,  which is a binary variable taking level 1 if each $M_k=1$, $k \in D$, and level 0 otherwise. The  multivariate log-linear regression model is:
\begin{eqnarray}\label{eq:1m}
\log P(\Mobs_D=1 \mid W,\XX)&=&  \beta^D_\emptyset + \beta^D_W W + \beta_X'^D \XX, \quad D \subseteq \{1, \ldots, K\}\\
\log P(\Yobs=1\mid \MMobs,W,\XX)& =&  \alpha_\emptyset  + \alpha_W W + \alpha^\tr_{M} \MM^{\mathrm{obs}} + \alpha_X^\tr \XX \label{eq:2m},
\end{eqnarray}
Under Assumptions \ref{ass:1} and \ref{ass:2}  and correct specification of the regression models in \eqref{eq:1m} and \eqref{eq:2m}, the  natural direct and indirect relative risks are   given by 
\begin{eqnarray}
RR^{\; dir}_{Y \mid \xx}&=& e^{\alpha_W} \notag\\\label{eq: rr-dir-mult}
RR^{\; ind}_{Y\mid \xx}&=& \dfrac{\sum_{\mm \in \{0,1\}^{K}} RR^{\; reg}_{Y|\mm.w\xx}\times P(\MMobs = \mm \mid W=1,\XX=\xx)}{\sum_{\mm \in \{0,1\}^{K}} RR^{\; reg}_{Y|\mm.w\xx}\times P(\MMobs = \mm \mid W=0,\XX=\xx)}. \notag\label{eq:rr-indir-mult}
\end{eqnarray}
See Appendix B for the proof, which relies on the mediation formula. Exposure-mediator and mediator-mediator interactions can be incorporated by adding the corresponding product terms, such as \(WM_k^{\mathrm{obs}}\) and \(M_D^{\mathrm{obs}}\) with \(|D|>1\), to the outcome model in \eqref{eq:2m}. The resulting natural direct and indirect effects are derived in Section B of the Supplementary Material.

The collection of probabilities \(P(M_D^{\mathrm{obs}}=1\mid W,\XX)\), for all nonempty subsets \(D\subseteq\mathcal K\), characterizes the joint distribution of the  mediators. The higher-order probabilities corresponding to subsets with \(|D|>1\) capture joint occurrence patterns among mediators and then reflect their dependence structure in the expression of the indirect effect. Even though the outcome model in \eqref{eq:2m} includes only the main effects of the mediators, the indirect effect is evaluated by averaging over their joint distribution. Parsimonious specifications of the  outcome model can be obtained by restricting the maximum order of the  product terms.

\subsection{Multiple Continuous Mediators}
\label{s:multipl_conmed}
Now we consider a set of continuous Gaussian mediators: $\MM(w) \sim MN(E[\MM(w)\mid \XX], \Sigma_{\MM(w)})$.
The model is
\begin{eqnarray}\label{eq:1mc}
E[\MMobs\mid W,\XX]&=&  \gamma_\emptyset + \gamma_W W + \gamma_X \XX
\end{eqnarray}
where $\gamma_{\emptyset}=(\gamma^k_{\emptyset})_{k \in K}$, $\gamma_{W}=(\gamma^k_{W})_{k \in K}$ and $\gamma_{\XX}$ is a $K \times |\XX|$ matrix with raw vector $(\gamma^{k}_{\XX})^\tr$, $k \in \{1, \ldots, K\}$ and 
\begin{eqnarray} 
\log P(\Yobs=1\mid W,\MMobs,\XX)& =&  \alpha_\emptyset  + \alpha_W W + \alpha^\tr_{M} \MMobs + \alpha_X^\tr \XX. \label{eq:2mc}
\end{eqnarray}

Under Assumptions \ref{ass:1} and \ref{ass:2} and correct specification of the regression models in \eqref{eq:1mc} and \eqref{eq:2mc}, the  natural direct and indirect relative risks are   given by 
\begin{eqnarray*}
RR^{\; dir}_{Y \mid \xx}&=& e^{\alpha_W},\\
RR^{\; ind}_{Y\mid \xx}&=&e^{\alpha^\tr_M \gamma_W}\dfrac{\exp\left\{\frac12 \alpha_M^\tr \Sigma_{\MM(1)}\alpha_M \right\}}{\exp\left\{\frac12 \alpha_M^\tr \Sigma_{\MM(0)}\alpha_M \right\}}.
\end{eqnarray*}
The derivation, based on the mediation formula, and extensions allowing for exposure--mediator interactions in model \eqref{eq:2mc} are presented in Section B of the Supplementary Material.

Similarly, mediator-mediator interactions can be included in Equation \eqref{eq:2mc} as follows:
\begin{eqnarray*} 
\log P(\Yobs=1\mid W,\MMobs,\XX)& =&  \alpha_\emptyset  + \alpha_W W + \alpha^\tr_{M} \MMobs + \frac12 \MM^{\tr obs}B \MMobs + \alpha_X^\tr \XX. 
\end{eqnarray*}
where \(B\) is the matrix containing the interactions' coefficients. In the simple case of two continuous mediators, B is
\[B =
\begin{pmatrix}
0 & \alpha_{12} \\
\alpha_{12} & 0
\end{pmatrix}.\]
 Under Assumptions \ref{ass:1} and \ref{ass:2} and  under correct specification of the regression models, the mediation formula implies that
\begin{eqnarray*}
    \lefteqn{P\left(Y(w, \MM(w^\ast)) =1 \mid \XX=\xx\right) }\\
    &=&
    \exp\left\{\alpha_\emptyset + \alpha_W w + \alpha_X^\tr \xx\right\} E\left[\exp\left\{\alpha_M^\tr \MMobs + \frac{1}{2}\MM^{'obs} B \MMobs \right\} \mid  W=w^\ast, \XX=\xx\right]
\end{eqnarray*}
where $E\left[\exp\left\{\alpha_M^\tr \MMobs + \frac{1}{2}\MM^{\tr obs} B \MMobs \right\} \mid  W=w^\ast, \XX=\xx\right]$ is the moment generating function of a quadratic form in the multivariate normal random vector $\MMobs \mid W=w^\ast, \XX=\xx$, with mean vector $\gamma_\emptyset + \gamma_W w^\ast + \gamma_X^{\tr} \xx$ and variance-covariance matrix $\Sigma_{\MM(w^\ast)}$, evaluated at linear coefficient $\alpha_M$ and quadratic coefficient matrix $B$ \citep{scarowsky1973quadratic}.
See Section B of the Supplementary Material for details and the resulting natural direct and indirect effects.

\subsection{Multiple Mixed Mediators}
\label{s:multiple_mixmed}
We next consider a setting with mixed mediators, comprising both continuous and binary variables. For simplicity we focus on settings with two mediators,  one continuous and one binary. Let $\MM(w) =(M_1(w), M_2(w))^\tr$ a random vector where  $M_2(w)$ is a binary variable taking values in $\{0,1\}$, and $M_1(w)$ is continuous. We assume that $M_1(w)|M_2(w) \sim N(E[M_1(w) \mid M_2(w),\XX], \sigma_{1|2}^2)$ for  $w \in \{0,1\}$.
We specify the following recursive regression models for the observed mediators and outcome:
\begin{eqnarray} \label{eq1:mix}
	\log P(\Mobs_2=1|W,\XX) &=& \delta_{\emptyset} + \delta_W W + \delta_X^\tr \XX,\\
	E[\Mobs_1|\Mobs_2,W,\XX]&=& \gamma_{\emptyset}+ \gamma_W W+ \gamma_{2} \Mobs_2 + \gamma_{W2}W\cdot \Mobs_2 + \gamma_X^\tr \XX \label{eq2:mix}\\
	\log(\Yobs=1|M_1,M_2, W, \XX) &=& \alpha_{\emptyset} + \alpha_W W + \alpha_1 \Mobs_1 + \alpha_2 \Mobs_2 + \alpha_X^\tr \XX \label{eq3:mix}
\end{eqnarray}

In Section B of the Supplementary Material, we show that under Assumptions \ref{ass:1} and \ref{ass:2} and  under correct specification of the regression models in \eqref{eq1:mix}-\eqref{eq3:mix}, the mediation formula yields the following expressions for the natural direct and indirect relative risks: 
\begin{eqnarray*}
   RR^{\;dir}_{Y\mid \xx} &=& \exp\{\alpha_W\}\\
    RR^{\;ind}_{Y\mid \xx},
    &=& \exp\{\alpha_1\gamma_W\}\times \dfrac{1- \exp\{\delta_{\emptyset}+\delta_W+\delta_X^\tr \xx\}\times (1+\exp\{\alpha_2+\alpha_1[\gamma_2+\gamma_{W2}]\})}{1- \exp\{\delta_{\emptyset}+\delta_X^\tr \xx\}\times (1+\exp\{\alpha_2+\alpha_1\gamma_2\})}.
\end{eqnarray*}
Exposure-mediator and mediator-mediator interactions can be incorporated by adding the corresponding product terms to the outcome model in \eqref{eq3:mix}. The resulting natural direct and indirect effects are derived in Section B of the Supplementary Material.

\section{Inference}
\label{s:inference}

The proposed regression-based approach defines a class of models which belongs to the curved exponential family both in case of binary and continuous mediators. While the relative risk measures offer highly desirable properties, inference for the related models presents non-trivial complexities.  Particularly when the mediators are binary,  the parameter space is subject to complex non-linear constraints, introducing  substantial  issues for inference and increasing the risk of out-of-range parameter estimates.  
A likelihood inference procedure is designed and  implemented by a separate maximization of the log-likelihood function for the outcome model and of the log-likelihood function for the  joint distribution of the mediators. 
Once maximum likelihood estimates (MLEs) are achieved for regression parameters $\underline{\alpha}$ and $\underline{\beta}$, MLEs for the direct and indirect effects are obtained by using the equivariance principle and  standard errors can be derived using the Delta method or bootstrap. We derive closed-form expressions for the standard errors using the Delta method (see Section C of the Supplementary Material).

\subsection{Model for the outcome}
Consider a conditional Bernoulli distribution for $\Yobs_i \mid (\underline{Z}_i=\underline{z}_i)  \sim Bern(\pi_i) $, where $\underline{Z}_i =   (W_i, \MMobs_i, \underline{\XX_i})'$ denotes the explanatory variable set  for the  outcome observed over a sample of size $n$, with
  $i=1,\dots, n$.  MLEs  are carried out by maximizing the log-likelihood function 
\begin{eqnarray*}
	\ell(\underline{\alpha};\underline{y}^{\text{obs}}, \underline{z})&=& \sum_{i=1}^n [ \yobs_i \log \pi_i + (1-\yobs_i) \log (1-\pi_i)], \\
	 &=&\sum_{i=1}^n [ \yobs_i (\underline{z}_i' \underline{\alpha}) + (1-\yobs_i) \log (1-\exp(\underline{z}_i' \underline{\alpha}))],
\end{eqnarray*}
where  $\pi_i = \exp(\underline{z}_i' \underline{\alpha})$. The score and the Hessian matrix are, respectively, 
\begin{equation*}
	u(\underline{\alpha}) = \frac{\partial \ell(\underline{\alpha})}{\partial \underline{\alpha}}= 
    \sum_{i=1}^n \bigg( \frac{\yobs_i -\exp(\underline{z}_i' \underline{\alpha})}{1-\exp(\underline{z}_i' \underline{\alpha})} \bigg)\underline{z}_i, \quad
    H(\underline{\alpha}) = \frac{\partial^2 \ell(\underline{\alpha})}{\partial \underline{\alpha} \partial \underline{\alpha}'}=- \sum_{i=1}^n \frac{\exp(\underline{z}_i' \underline{\alpha})(1-\yobs_i)}{(1-\exp(\underline{z}_i' \underline{\alpha}))^2} \underline{z}_i \underline{z}_i'.
\end{equation*}
As the likelihood equations do not provide a closed-form solution, an  iterative procedure 
is required for parameter estimation. Notably, the choice of a log-link introduces a parameter space constraint since   both  $\log (\pi_i)$  and the linear predictor  $\underline{z}_i' \underline{\alpha} $ are negative for any $i =1,\dots, n$, unlike the canonical  logit-link where the linear predictor  belongs to the entire real space. Constrained MLEs can be numerically obtained, for instance, by using  the \texttt{glm2} package in R for binomial family with log-link. 
We recommend using \texttt{glm2} rather than \texttt{glm}, since the latter may fail to converge or produce out-of-range estimates;
\texttt{logbin} \citep{donoghoe2018logbin}  is another package available in R for fitting log-link univariate regression models.

\subsection{Model for multiple binary mediators}\label{sec:inference-bin-med}

The joint model of  $K$ binary mediators $\MMobs_i|(\underline{S_i} =\underline{s_i})$, for $i=1,\dots,n$, follows a multivariate Bernoulli distribution with probability parameter vector $\underline{\lambda_i}=(\lambda_{D,i})_{D \subseteq K}$, where $\underline{S_i}=(W_i,\XX_i)'$ is the set of explanatory variables  and
the generic element of $\underline{\lambda_i}$ is the joint probability $\lambda_{D,i}=P(\MMobs_{D,i}=\underline{1}, \MMobs_{K \setminus D, i}=\underline{0}| \underline{S_i}= \underline{s_i})$, for any $D \subseteq K$.  The mean parameter vector $\underline{\mu_i}=(\mu_{D,i})_{D \subseteq K}$ is a collection of marginal probabilities $\mu_{D,i} =P(\MMobs_{D,i}=\underline{1}|\underline{S_i}=\underline{s_i})$. The mean and the probability parameters are linked by the one-to-one M\"obius inversion that, expressed in matrix form, is
$\underline{\lambda_i} = \mathbb{M}  \underline{\mu_i}$  and $ \underline{\mu_i} = \mathbb{Z}  \underline{\lambda_i}$, 
where $\mathbb{M}$ is a square symmetric matrix, known as M\"obius matrix, and  $\mathbb{Z}=\mathbb{M}^{-1}$ \citep{lup-rov:2017}. 
We adopt a  generalized linear regression model based on the log-mean function  such that
\begin{equation}\label{eq:loglink}
    \underline{\eta_{i}} = \log \underline{\mu_{i}}= \underline{s_i}' \underline{\beta}, \qquad i =1,\dots,n
\end{equation}
where $\underline{\beta}$ belongs to a constrained parameter space as  the vector $\underline{\eta_{i}}$ is negative and $1'\underline{\lambda_i}=1$. In order to avoid out-of range estimates during the maximization procedure, the log-likelihood is conveniently written by exploiting the one-to-one mapping between $\log \underline{\mu_{i}}$ and   the canonical log-linear parameter $\underline{\theta_i}= \mathbb{M}' \log(\underline{\lambda_i})$ that is defined in a rectangular  space. More specifically, the mapping, based on a recursive application of the M\"obius inversion,  is
\begin{equation*}
    \log \underline{\mu_i}= \underline{s_i}' \underline{\beta}= \log \mathbb{Z}\exp (\mathbb{Z}'\underline{\theta_i}), \qquad i =1,\dots,n,
\end{equation*}
and the resulting log-likelihood function is 
\begin{equation}\label{eq:lik_individual}
    \ell(\underline{\beta}; \underline{y}^{\text{obs}},\underline{s})= \sum_i \yobs_i \mathbb{Z}' \underline{\theta_i} - \log [1' \exp(\mathbb{Z}'\underline{\theta_i})].
\end{equation} 
We implemented an algorithm for MLEs that leverages  the maximization procedure proposed by  \citet{eva-for:2013} for a  broad class of link functions including the log-mean link in \eqref{eq:loglink} as special case. Interestingly, the  link we propose  simplifies the fitting procedure as it allows for closed-form parameter transformations via M\"{o}bius inversion, eliminating the need for additional iterative procedures when switching between parameterizations within the estimation algorithm. Conversely, a  crucial issue arises in defining a starting value of $\underline{\beta}$  that satisfies the constraints of the parameter space for the linear predictor. We propose using the  estimates derived from the  independence model for the mediators, which serves as a valuable and efficient starting point. Further technical details on the maximization procedure are deferred to Section D of the Supplementary Material, while an algorithm implemented in R is available at \url{https://github.com/ariannanuti/mediation_project}. 

The log-likelihood in \eqref{eq:lik_individual} is written for a set of individual data that is typically available when some of the explanatory variables $\XX$ are continuous. In case the variables should be all binary (or categorical) and data are collected in form of counts of a contingency table,  MLEs can be obtained by maximizing the log-likelihood function of a Multinomial distribution for the joint distribution of $(\MMobs, \underline{S})'$. Optimization procedures for likelihood inference of multivariate generalized  regression models for  categorical  data  can be performed, for instance, by using the \texttt{cta} package in R. These procedures are illustrated and discussed  in \citet{lup-mar-tar:2025} and in \citet{lupparelli:2019} more specifically for  log-mean regressions.

\subsection{Model for multiple continuous or mixed mediators}

When the $K$ mediators are continuous, we assume that $\MMobs_i|(\underline{S_i}=\underline{s_i}) \sim N (\underline{s_i}' \underline{\beta}, \Sigma)$. Likelihood inference for multivariate Gaussian regression model
$$
\MMobs_i = \underline{s_i}' \underline{\beta} + \underline{\epsilon_i},  \qquad E(\underline{\epsilon_i}) =0, \;\; Var( \underline{\epsilon_i})= \Sigma
$$
can be performed by using well-established algorithms that provide MLEs for regression coefficients $\underline{\beta}$ and for the variance-covariance matrix $\Sigma$ modelling the dependence of the error term; for instance maximization procedures in \texttt{lavaan} or \texttt{systemfit} packages of R.

When  $\MMobs=(\MMobs_1, \MMobs_2)'$ is mixed, with $\MMobs_1$ a vector of continuous variables  and $\MMobs_2$  a vector of binary variables,  we factorize the conditional joint distribution as  $P(\MMobs|\underline{S})=P(\MMobs_1|\underline{S}, \MMobs_2)P(\MMobs_2| \underline{S})$ where $\MMobs_1\mid\underline{S},\MMobs_2$ is assumed to follow a multivariate Gaussian distribution and $\MMobs_2\mid\underline{S}$ a multivariate Bernoulli distribution.
We perform MLEs for the marginal model  $\MMobs_2| \underline{S}$ of binary mediators using the likelihood inference described in Section \ref{sec:inference-bin-med} and then we fit the conditional model  $\MMobs_1|(\underline{S}, \MMobs_2)$ of the Gaussian mediators.

\section{Applications}
\label{s:applicantions}

We illustrate the proposed methodology through two empirical applications involving different mediator types and substantive settings.

For each study, we show results under four specifications of the mediator model: a joint model for the conditional distribution of all mediators, a conditionally independent mediator model, and two separate single-mediator analyses. The joint model is treated as the primary specification, whereas the remaining analyses assess the effect of simplifying the dependence structure among the mediators. All models are adjusted for baseline covariates.

\subsection{Continuous Mediators: Physical Activity and Diabetes Risk}
\label{s:application_continuous}

The proposed methodology is applied to evaluate the direct effect of physical activity on diabetes risk and the indirect effect operating through average systolic blood pressure and total cholesterol, using data from the National Health and Nutrition Examination Survey (NHANES) available in the \texttt{NHANES} R package \citep{Pruim2015}. NHANES is a health and nutrition survey conducted by the US National Center for Health Statistics on the non-institutionalized civilian population of the United States. The R package dataset contains 10,000 observations resampled from the 2009--2010 and 2011--2012 survey cycles to account for the oversampling of specific population groups. We restrict the analysis to participants older than 50 years, yielding a final sample of $N=2721$ subjects.

For each subject, we observe a binary treatment variable, \(W_i\), indicating whether the participant reported moderate or vigorous-intensity sports, fitness, or recreational activities; a binary outcome, \(\Yobs_i\), indicating whether the participant reported a diabetes diagnosis by a doctor or health professional; two continuous mediators, \(\Mobs_{i1}\), average systolic blood pressure, and \(\Mobs_{i2}\), total cholesterol; and ($p=2$) covariates, \(X_{i1}\), gender, and \(X_{i2}\), age in years.

Under Assumptions~\ref{ass:1} and~\ref{ass:2}, we estimate marginal natural direct, indirect, and total effects on the risk ratio scale. For the single-mediator analyses, we use the 
mediator and outcome models in Equations~(\ref{eq:3}) and~(\ref{eq:4}). For the multiple-mediator analysis, we use the joint framework described in Section~\ref{s:multipl_conmed} with details reported in Section B of the Supplementary Material.
Specifically, systolic blood pressure and total cholesterol are modeled through a bivariate normal distribution, allowing for conditional association, and the outcome model includes both mediators and their interaction. All models are adjusted for gender and age.

Table~\ref{tab:marg_effects_cont} reports the estimated effects under four mediator specifications: the joint bivariate normal model, the conditionally independent mediator model, and two separate single-mediator analyses.

\begin{table}[ht]
\centering
\label{tab:marg_effects_cont}
\begin{tabular}{llrrrr}
\toprule
Mediator specification & Effect & Estimate & SE & CI lower & CI upper \\ 
\midrule
\multicolumn{6}{l}{\textit{Systolic Blood Pressure and total cholesterol: joint model}} \\ 
 & Direct & 0.759 & 0.063 & 0.645 & 0.894 \\ 
 & Indirect & 0.930 & 0.017 & 0.897 & 0.965 \\ 
 & Total & 0.706 & 0.060 & 0.598 & 0.833 \\ 
\addlinespace
\multicolumn{6}{l}{\textit{Systolic Blood Pressure and total cholesterol: independent model}} \\ 
 & Direct & 0.759 & 0.063 & 0.645 & 0.894 \\ 
 & Indirect & 0.931 & 0.018 & 0.897 & 0.966 \\ 
 & Total & 0.706 & 0.060 & 0.598 & 0.834 \\
\addlinespace
\multicolumn{6}{l}{\textit{Systolic Blood Pressure only}} \\ 
 & Direct & 0.698 & 0.060 & 0.590 & 0.825 \\ 
 & Indirect & 1.000 & 0.004 & 0.991 & 1.009 \\ 
 & Total & 0.698 & 0.060 & 0.590 & 0.825 \\ 
\addlinespace
\multicolumn{6}{l}{\textit{Total cholesterol only}} \\ 
 & Direct & 0.747 & 0.062 & 0.634 & 0.880 \\ 
 & Indirect & 0.944 & 0.017 & 0.912 & 0.977 \\ 
 & Total & 0.705 & 0.060 & 0.597 & 0.832 \\ 
\bottomrule
\end{tabular}
\caption{Marginal natural direct, indirect, and total effects of physical activity on the risk of diabetes, on the risk ratio scale. The continuous mediators are average systolic blood pressure and total cholesterol. 
Results are reported under the joint mediator model, the conditionally independent mediator model, and the two separate single-mediator analyses. Standard errors and \(95\%\) confidence intervals were obtained using the delta method.
}
\end{table}

Across all specifications, physical activity is associated with a statistically significant reduction in diabetes risk, with total-effect estimates close to \(0.70\) $(se=0.06)$. Thus, the estimated total effect suggests that physical activity reduces the risk of diabetes by approximately 30\%.

Under the joint mediator model, the direct and indirect effects are \(0.759\) $(se=0.063)$  \(0.930\), $(se=0.017)$, 
respectively, and both are statistically significant.
These findings suggest that the protective effect of physical activity on diabetes risk is predominantly direct, with a smaller but statistically significant proportion mediated through systolic blood pressure and total cholesterol.
The joint and conditionally independent mediator models yield virtually identical estimates, suggesting that residual conditional association between the two mediators has negligible impact in this application. This is consistent with the lack of evidence for a mediator--mediator interaction in the outcome model, although the joint model remains the more general specification. The single-mediator analyses show that systolic blood pressure alone has an essentially null and non-significant indirect effect, whereas total cholesterol alone has a statistically significant indirect effect of \(0.944\) $(se=0.017)$. Thus, the mediated component is mainly driven by total cholesterol, but the joint indirect effect is slightly farther from the null than the cholesterol-specific effect, suggesting that the joint analysis captures a broader mediated pathway than either single-mediator analysis alone.

\subsection{Binary Mediators: Anxious Attachment and Unwanted Pursuit Behaviors}
\label{s:application_binary}

The proposed methodology is applied to evaluate the direct effect of anxious attachment before relationship breakup on subsequent unwanted pursuit behavior (UPB) and the joint indirect effect operating through negative affectivity and breakup initiator status. We use the \texttt{UPBdata} dataset from the \texttt{medflex} R package, previously analysed by \citet{loeys2013flexible}. 
We retain the same treatment (attachment anxiety during the relationship), primary mediator (negative affect experienced during the breakup), and outcome (post-breakup UPB), and extend their analysis by including breakup initiator status as a second mediator.  
Attachment anxiety is measured before the breakup and may influence who initiates the breakup  \citep{loeys2013flexible}.  At the same time, initiator status may mediate part of the effect of attachment anxiety on UPB.

The data come from the Interdisciplinary Project for the Optimization of Separation trajectories (IPOS), a large-scale study of divorce and separation in Flanders. The project recruited individuals who divorced between March 2008 and March 2009 through four major Flemish courts. The dataset includes 385 individuals who completed questionnaires on romantic relationships, breakup characteristics, and post-breakup behavior. 

For each unit, we observe a binary treatment \(W_i\), indicating whether anxious attachment before the breakup was above the sample mean; a binary outcome \(\Yobs_i\), indicating whether the individual reported UPB toward the ex-partner after the breakup; two binary mediators, \(\Mobs_{i1}\),
indicating whether negative affectivity, dichotomized at the sample median, was high, and \(\Mobs_{i2}\), indicating whether the separation was initiated by the respondent's ex-partner; and baseline covariates \(\XX_i\), comprising gender, age, and education level, categorized as low, intermediate, or high.

Under Assumptions~\ref{ass:1} and~\ref{ass:2}, we estimate marginal natural direct, indirect, and total effects on the risk ratio scale. For the single-mediator analyses, we use the binary-mediator models in Equations~(\ref{eq:1}) and~(\ref{eq:2}); for the multiple-mediator analyses, we use the framework in Equations~(\ref{eq:1m}) and~(\ref{eq:2m}), with details reported in Section B of the Supplementary Material. 

Table~\ref{tab:marg_effects_binary} reports the estimated marginal natural direct, indirect, and total effects under the four specifications for the mediators' models.

\begin{table}[ht]
\centering
\label{tab:marg_effects_binary}
\begin{tabular}{llrrrr}
\toprule
Mediator specification & Effect & Estimate & SE & CI lower & CI upper \\ 
\midrule
\multicolumn{6}{l}{\textit{Negative affectivity and ex-partner initiation: joint model}} \\ 
 & Direct & 1.372 & 0.185 & 1.053 & 1.788 \\ 
 & Indirect & 1.217 & 0.062 & 1.102 & 1.345 \\ 
 & Total & 1.671 & 0.220 & 1.290 & 2.163 \\ 
\addlinespace
\multicolumn{6}{l}{\textit{Negative affectivity and ex-partner initiation: independent model}} \\ 
 & Direct & 1.372 & 0.185 & 1.053 & 1.788 \\ 
 & Indirect & 1.212 & 0.058 & 1.104 & 1.330 \\ 
 & Total & 1.663 & 0.218 & 1.286 & 2.150 \\ 
\addlinespace
\multicolumn{6}{l}{\textit{Negative affectivity only}} \\ 
 & Direct & 1.459 & 0.200 & 1.115 & 1.908 \\ 
 & Indirect & 1.145 & 0.049 & 1.053 & 1.245 \\ 
 & Total & 1.670 & 0.228 & 1.278 & 2.182 \\  
\addlinespace
\multicolumn{6}{l}{\textit{Ex-partner initiation only}} \\ 
 & Direct & 1.522 & 0.208 & 1.164 & 1.991 \\ 
 & Indirect & 1.101 & 0.039 & 1.026 & 1.180 \\ 
 & Total & 1.675 & 0.228 & 1.284 & 2.187 \\ 
\bottomrule
\end{tabular}
\caption{Marginal natural direct, indirect, and total effects of high anxious attachment on the risk of reporting unwanted pursuit behavior (UPB), on the risk ratio scale. The binary mediators are high negative affectivity and ex-partner-initiated divorce. 
Results are reported under the joint mediator model, the conditionally independent mediator model, and the two separate single-mediator analyses. Standard errors and \(95\%\) confidence intervals were obtained using the multivariate delta method.
}
\end{table}

Across all specifications, high anxious attachment is associated with a significantly higher risk of reporting unwanted pursuit behavior. The estimated total effect is highly stable, ranging from \(1.663\) to \(1.675\), with all confidence intervals excluding the null value. 

The joint mediator model leads to  statistically significant direct and indirect effects of \(1.372\) $(se=0.185)$ and \(1.217\) $(se=0.062)$, respectively.
Under the joint mediator model, the estimated total effect is \(1.671\), with statistically significant direct and indirect effects of \(1.372\) and \(1.217\), respectively. When the mediators are considered separately, the indirect effects are smaller: \(1.145\)  $(se=0.049)$ for negative affectivity and \(1.101\) $(se=0.039)$ for ex-partner initiation, although still statistically significant. Correspondingly, the direct effects increase to \(1.459\) $(se=0.200)$ and \(1.522\) $(se=0.208)$. 
Thus, both mediators appear to channel part of the effect of high attachment anxiety. This mediated component can be disentangled in the multiple-mediator analysis, whereas it is absorbed into the estimated direct effect in the single-mediator analyses.

The joint and conditionally independent specifications produce similar results. Hence, in this application, accounting for the conditional dependence between negative affectivity and ex-partner initiation has little influence on the estimated marginal effects. Nevertheless, the joint model remains the preferred specification because it directly estimates a coherent joint distribution for the two binary mediators, without imposing conditional independence.

\section{Discussion}
\label{s:discuss}
We introduce a novel recursive regression framework for  mediation analysis with multiple contemporaneous mediators that accommodates binary, continuous, and mixed mediators in a unified setting, and naturally incorporates exposure-mediator and mediator-mediator interactions.  The latter comes directly from the joint modelling of the mediators' distribution, regardless of its continuous or discrete nature. By defining causal estimands on the relative risk scale, our approach overcomes the non-collapsibility of traditional odds-ratio-based frameworks and provides a non-approximated parametric decomposition of the total effect into natural direct and joint indirect relative risks. This enhances causal interpretation and allows for exact maximum likelihood estimation of the causal risk-ratios  enabling the Delta method for standard errors.
From a computational standpoint, while the choice of a log-mean link for multivariate binary variables introduces non-trivial parameter space constraints, we have implemented a maximum likelihood estimation routine that is numerically stable and efficient. Nevertheless, further inferential frameworks could be explored to enhance efficiency in non-standard scenarios, such as very large sample sizes or the presence of rare binary mediators.

The proposed framework can be applied to  mediation analysis in both observational studies and randomized trials. However, the use of relative-risk-based models remains limited in case-control studies when selection is outcome-dependent. 
A promising direction for future research involves adapting our approach under covariate-based selection or developing appropriate bias-correction methods for outcome-dependent sampling.

\section*{Supplementary Material}

The Supplementary Material is available from the authors upon request.

\bibliographystyle{agsm}
\bibliography{bibliography}

\end{document}